\documentclass[aps,prb,amsmath,amssymb]{revtex4}
\usepackage{graphicx}
\usepackage{bm}
\usepackage{bbold}

\def\bH{ \bm{H} }

\def\b0{ \bm{0} }
\def\bhx{ \hat{\bm{x}} }

\def\bhz{ \hat{\bm{z}} }

\def\sgn{\, \mathrm{sgn}\, }

\def\diag{\,\mathrm{diag}\,}

\begin{document}
\title{Majorana modes in multiband superconducting quantum wires}

\author{K. V. Samokhin\footnote{e-mail: kirill.samokhin@brocku.ca}}

\affiliation{Department of Physics, Brock University, St. Catharines, Ontario L2S 3A1, Canada}
\date{\today}

\begin{abstract}
We calculate analytically the spectrum of the Andreev bound states in a half-infinite superconducting wire with an arbitrary number of bands crossing the chemical potential. The normal state of the wire is assumed to have 
an antiunitary symmetry ${\cal A}$ (time reversal or its combination with a crystallographic point group operation), with ${\cal A}^2=-1$ or $+1$. This symmetry may be broken by 
the superconducting order parameter and/or the boundary scattering. We present a model-independent proof of the existence of one Majorana mode near the end of the wire with an odd number of bands.
\end{abstract}

\maketitle

\section{Introduction}
\label{sec: intro}

One of the most promising experimental setups for the realization of Majorana fermions (MFs) in condensed matter\cite{Alicea12,LF12,Bee13} is based on Kitaev's lattice model of a one-dimensional 
(1D) spin-polarized $p$-wave superconductor.\cite{Kit01} In the topologically nontrivial state this model has unpaired, or ``dangling'', zero-energy Andreev boundary states (ABSs), interpreted as MFs. 
One can engineer a Kitaev chain-like system using a semiconducting wire in which superconductivity is induced by proximity with a conventional bulk superconductor.\cite{Lutchyn10,Oreg10,InSb-wire}

The crucial ingredients for the quantum-wire MF proposals are an antisymmetric spin-orbit coupling (SOC), see Ref. \onlinecite{Rashba-model}, and time reversal (TR) symmetry breaking, 
\textit{e.g.}, by an external magnetic field. The former requires the absence of an inversion center in the system,\cite{NCSC-book} which naturally occurs in a quantum wire placed on a substrate. 
The antisymmetric SOC lifts the spin degeneracy of the electron states almost everywhere in the 1D Brillouin zone, except the TR-invariant wave vectors, 
producing an even number $N$ of nondegenerate 1D Bloch bands crossing the chemical potential. 
A nonzero magnetic field is required to remove the band degeneracies at the TR-invariant points and make odd values of $N$, in particular, $N=1$, possible. 
The magnetic field has another, perhaps less appreciated, role in that it endows the wire with an antiunitary symmetry, which is crucial for the proper definition of the gap function and 
the construction of the Bogoliubov-de Gennes (BdG) Hamiltonian.\cite{Sam17-AoP}

It was argued in Ref. \onlinecite{Kit01}, see also Refs. \onlinecite{PL10} and \onlinecite{LSDS11}, that an odd number of superconducting bands is required for the MFs to be topologically stable. 
However, to the best of the author's knowledge, an explicit analytical proof of this statement for a half-infinite multiband wire is still lacking. 
The goal of our paper is to fill this gap and calculate the ABS spectrum in a quasi-1D superconducting wire with $N$ bands, for an arbitrary symmetry of the boundary scattering and arbitrary complex
values of the $N$-component order parameter. 
Using rather general symmetry arguments and the semiclassical (Andreev) approach allows us to calculate the ABS spectrum independently 
of any particular model for the electronic band structure. Throughout the paper we use the units in which $\hbar=1$.

\section{Symmetry analysis}
\label{sec: symmetry}

We consider a clean quasi-1D wire on a flat substrate. The substrate occupies the $z<0$ half-space and the wire is oriented along the $x$ axis. 
The potential energy $U(x,y,z)$ affecting the electrons in the wire is assumed to be periodic in $x$, with the period $d$, and confining in both $y$ and $z$ directions. 
Since the substrate breaks the reflection symmetry $\sigma_z$, the system lacks an inversion center but may still be invariant under the mirror reflections $\sigma_x$ and/or $\sigma_y$. 
Therefore, in the absence of TR symmetry breaking the symmetry of the wire is described by one of the five quasi-1D point groups 
$\mathbb{G}$: $\mathbf{C}_1=\{E\}$, $\mathbf{D}_x=\{E,\sigma_x\}$, $\mathbf{D}_y=\{E,\sigma_y\}$, $\mathbf{C}_2=\{E,\sigma_x\sigma_y\}$, and $\mathbf{V}=\{E,\sigma_x,\sigma_y,\sigma_x\sigma_y\}$, see Ref. \onlinecite{Sam17-PRB}.

In the presence of an external magnetic field $\bH$, one has to use the magnetic point groups, or the magnetic classes, $\mathbb{G}_M$,
whose elements leave both the scalar potential $U(x,y,z)$ and the pseudovector $\bH$ invariant.
There are three types of the magnetic classes.\cite{LL-8} Type I classes $\mathbb{G}_M=\mathbb{G}$ do not involve the TR operation $K$ at all, neither by itself nor in combination 
with $\sigma_x$ and $\sigma_y$. Type II classes contain $K$ and, therefore, describe TR invariant systems, with $\bH=\b0$. These classes are obtained from the ordinary point groups as $\mathbb{G}_M=\mathbb{G}+K\mathbb{G}$.
Type III magnetic classes contain the TR operation only in combination with $\sigma_x$ and $\sigma_y$. In the notation for these classes, 
$\mathbb{G}_M=\mathbb{G}(\tilde{\mathbb{G}})$, the unitary subgroup $\tilde{\mathbb{G}}$ comprises the elements of $\mathbb{G}$ which are not multiplied by $K$. In this paper, we focus on Type II and Type III quasi-1D 
superconductors, which correspond to multiband $s$-wave and $p$-wave wires, respectively, see below.

The normal state of an infinite wire is described by the set of 1D electron Bloch bands $\xi_n(k)$, with the band energies counted from the chemical potential, and the corresponding Bloch spinor states $|k,n\rangle$, 
which are labelled by the band index $n$ and the wave vector $k$. Other than being periodic in the reciprocal space, $\xi_n(k+2\pi/d)=\xi_n(k)$, the bands do not have any additional symmetries, in general. 
However, in order for the bands to support the formation of the Cooper pairs with zero center-of-mass momentum, the band dispersions have to satisfy $\xi_n(k)=\xi_n(-k)$, 
\textit{i.e.}, the magnetic class has to contain at least one symmetry element which transforms $k$ into $-k$. 
While this is obviously the case for all Type II classes, in which the Cooper pairing occurs between time-reversed states, there are only three Type III classes that have the requisite 
$k\to-k$ symmetry:\cite{Sam17-AoP}
\begin{equation}
\label{Type-III-classes-1}
  \mathbf{D}_y(E)=\{E,K\sigma_y\},
\end{equation}
corresponding to $U(x,-y,z)=U(x,y,z)$ and $\bH=H_x\bhx+H_z\bhz$,   
\begin{equation}
\label{Type-III-classes-2}
  \mathbf{V}(\mathbf{D}_x)=\{E,\sigma_x,K\sigma_y,K\sigma_x\sigma_y\},
\end{equation}
corresponding to $U(-x,y,z)=U(x,-y,z)=U(x,y,z)$ and $\bH=H_x\bhx$, and
\begin{equation}
\label{Type-III-classes-3}
  \mathbf{V}(\mathbf{C}_2)=\{E,K\sigma_x,K\sigma_y,\sigma_x\sigma_y\},
\end{equation}
corresponding to $U(-x,y,z)=U(x,-y,z)=U(x,y,z)$ and $\bH=H_z\bhz$.
We see that any deviation of the magnetic field from the $xz$ plane produces asymmetric bands and is therefore detrimental for superconductivity. 
For symmetric bands, the ``Fermi surface'' is given by the set of $2N$ Fermi wave vectors $rk_{F,n}$, which are the roots of the equations $\xi_n(k)=0$. Here $r=\pm$, $n=1,...,N$, and $N$ is the number of nondegenerate bands 
crossing the Fermi level. While in Type II systems $N$ is even, in Type III systems $N$ can be even or odd. 

The spectrum of fermionic quasiparticles in the superconducting state is obtained by diagonalizing the BdG Hamiltonian, which is obtained as follows. We start with 
the general mean-field pairing Hamiltonian in some arbitrary basis of single-particle fermionic states $|i\rangle$:
\begin{equation}
\label{H-MF-general}
  \hat H=\sum_{ij}\varepsilon_{ij}\hat c_i^\dagger\hat c_j+\frac{1}{2}\sum_{ij}\bigl(\Delta_{ij}\hat c_i^\dagger\hat{\tilde c}_j^\dagger+\mathrm{H.c.}\bigr).
\end{equation}
Here $\varepsilon_{ij}=\langle i|\hat\varepsilon|j\rangle$ and the gap functions $\Delta_{ij}=\langle i|\hat\Delta|j\rangle$ are the matrix elements of the single-particle Hamiltonian $\hat\varepsilon$ and 
the gap operator $\hat\Delta$, respectively. In order to properly define the latter, one has to pair up fermions in the states related by an antiunitary operation ${\cal A}$ 
(Ref. \onlinecite{Sam17-AoP}). 
For each state $|i\rangle$, we introduce its ${\cal A}$-transformed counterpart ${\cal A}|i\rangle$, as well as the corresponding creation and annihilation operators  
$\hat{\tilde c}_i^\dagger={\cal A}\hat c_i^\dagger{\cal A}^{-1}$ and $\hat{\tilde c}_i={\cal A}\hat c_i{\cal A}^{-1}$. 
Then, the gap function $\Delta_{ij}$ is a measure of the pairing between the states $|i\rangle$ and ${\cal A}|j\rangle$. The BdG Hamiltonian corresponding to the model (\ref{H-MF-general}) has the following form: 
\begin{equation}
\label{H-BdG-general}
  \hat{\cal H}_{BdG}=\left(\begin{array}{cc}
                 \hat\varepsilon & \hat{\Delta} \\
		  \hat{\Delta}^\dagger & -\hat{\tilde\varepsilon}
                 \end{array}\right),
\end{equation}
where 
\begin{equation}
\label{hole-Hamiltonian}
  \hat{\tilde\varepsilon}={\cal A}^{-1}\hat\varepsilon{\cal A}.
\end{equation}
The operators and $\hat\varepsilon$ and $\hat{\tilde\varepsilon}$ can be interpreted as the Hamiltonians of electron-like and hole-like quasiparticles in the normal state. 
It should be noted though that there are no separate electron and hole Hilbert spaces, so that the first-quantization operators $\hat\varepsilon$, $\hat{\tilde\varepsilon}$, and $\hat\Delta$ 
all act in the same single-particle Hilbert space.

We assume that the ``bulk'' of our quasi-1D wire is invariant under the antiunitary operation ${\cal A}$, while the boundaries and/or the superconducting order parameter do not necessarily respect this symmetry. 
In other words, if the wire occupies the region 
$0\leq x\leq l$, then  inside this region $\hat\varepsilon$ commutes with ${\cal A}$ and $\hat{\tilde\varepsilon}=\hat\varepsilon$, but outside $\hat{\tilde\varepsilon}$ may be different from $\hat\varepsilon$.
While in Type II systems, one can use as ${\cal A}$ the TR operation $K$, which is the usual convention in theory of superconductivity,\cite{TR-pairing}
in Type III systems one should use ${\cal A}=K\sigma_y$, according to Eqs. (\ref{Type-III-classes-1}), (\ref{Type-III-classes-2}), and (\ref{Type-III-classes-3}).

In the bulk of the wire, we use the basis of the Bloch states $|i\rangle=|k,n\rangle$ in Eq. (\ref{H-MF-general}) and obtain: 
\begin{equation}
\label{H-MF-infinite-bulk}
  \hat H=\sum_{k,n}\xi_n(k)\hat c^\dagger_{kn}\hat c_{kn}+\frac{1}{2}\sum_{k,n}\bigl[\Delta_{n}(k)\hat c^\dagger_{kn}\hat{\tilde c}^\dagger_{kn}+\mathrm{H.c.}\bigr],
\end{equation}
where $\hat{\tilde c}^\dagger_{kn}={\cal A}\hat c^\dagger_{kn}{\cal A}^{-1}=t_n(k)\hat c^\dagger_{-k,n}$ and the phase factors $t_n$ are defined by the expression
\begin{equation}
\label{t_n-def}
  {\cal A}|k,n\rangle=t_n(k)|-k,n\rangle.
\end{equation}
The parity of $t_n$ and, therefore, that of the gap functions depends on whether ${\cal A}^2=-1$ or $+1$, when acting on spin-1/2 spinors. 
In Type II wires, we have ${\cal A}=K$, therefore, ${\cal A}^2=-1$ and $t_n(-k)=-t_n(k)$. In Type III wires, we have ${\cal A}=K\sigma_y$, therefore, ${\cal A}^2=1$ and 
$t_n(-k)=t_n(k)$. Accordingly, the gap functions in Type II wires correspond to a 1D ``$s$-wave'' pairing:
\begin{equation}
\label{Type-II-OP}
  \Delta_{n}(k)=\eta_n=\Delta_{n}(-k).
\end{equation}
In contrast, in Type III wires the gap functions correspond to 1D ``$p$-wave'' pairing:
\begin{equation}
\label{Type-III-OP}
  \Delta_{n}(k)=i\eta_n\varphi_n(k)=-\Delta_{n}(-k).
\end{equation}
Here $\eta_n$ are the superconducting order parameter components and $\varphi_n(k)$ are real odd functions, \textit{e.g.}, $\varphi_n(k)\propto k$. 
In both cases, the gap function phases are chosen so as to ensure that the action of the antiunitary operation on the 
order parameter components is equivalent to complex conjugation. While an ${\cal A}$-invariant superconducting state would correspond to real $\eta_n$, we do not make this assumption here and consider general complex $\eta_n$.

\section{ABS spectrum}
\label{sec: ABS}
 
In an infinite uniform superconducting wire, the energies of the Bogoliubov excitations are given by $E_n(k)=\sqrt{\xi_n^2(k)+|\Delta_n(k)|^2}$. While the bulk spectrum is gapped, there might exist subgap states 
localized near the ends. The energies of these states can be found using the semiclassical approach.\cite{And64} 
To make analytical progress, we assume that the order parameters $\eta_1,...,\eta_N$ in the wire do not depend on $x$. Then,
the semiclassical ``envelope'' wave function $\psi_{rk_{F,n}}(x)$ of the quasiparticles near the Fermi point $rk_{F,n}$ in the $n$th band, which varies slowly on the scale of $k_{F,n}^{-1}$, satisfies the Andreev equation:    
\begin{equation}
\label{And-eq-gen}
	\left(\begin{array}{cc}
		-iv_{n,r}\dfrac{d}{dx} & \Delta_n(rk_{F,n}) \\
		\Delta_n^*(rk_{F,n}) & iv_{n,r}\dfrac{d}{dx}
	\end{array}\right)\psi_{rk_{F,n}}(x)=E\psi_{rk_{F,n}}(x).
\end{equation}
Here $v_{n,r}=(\partial\xi_n/\partial k)|_{k=rk_{F,n}}$ is the quasiparticle group velocity near the Fermi point and $|v_{n,\pm}|=v_{F,n}$ is the Fermi velocity in the $n$th band. 

In a half-infinite ($x\geq 0$) wire, the solution of Eq. (\ref{And-eq-gen}) corresponding to the subgap ABS localized near $x=0$ has the form $\psi_{rk_{F,n}}(x)=\phi(rk_{F,n})e^{-\Omega_nx/v_{F,n}}$, where 
\begin{equation}
\label{Andreev amplitude}
	\phi(rk_{F,n})=\psi_{rk_{F,n}}(0)=C(rk_{F,n})\left(\begin{array}{c}
		\dfrac{\Delta_n(rk_{F,n})}{E-i\Omega_n\sgn v_{n,r}} \\ 1
	\end{array}\right),
\end{equation}
$\Omega_n=\sqrt{|\Delta_n(rk_{F,n})|^2-E^2}$, and $C$ is a coefficient. The semiclassical approximation breaks down near the end of the wire due to a rapid variation of the boundary potential, which causes 
transitions between the states corresponding to different Fermi wave vectors. This can be described by a phenomenological boundary condition for the envelope wave functions. 

The Fermi wave vectors are classified as either incident, for which $v_{n,r}<0$, or reflected, for which $v_{n,r}>0$. 
We denote the former $k^{\mathrm{in}}_{1},...,k^{\mathrm{in}}_N$ and the latter $k^{\mathrm{out}}_{1},...,k^{\mathrm{out}}_N$, with $k^{\mathrm{out}}_{n}=-k^{\mathrm{in}}_{n}$.
From Eq. (\ref{Andreev amplitude}), we obtain:
\begin{equation}
\label{phi-in-out}
  \phi(k^{\mathrm{in}}_n)=C(k^{\mathrm{in}}_n)\left(\begin{array}{c}
		\alpha^{\mathrm{in}}_n \\ 1
	\end{array}\right),\quad
  \phi(k^{\mathrm{out}}_n)=C(k^{\mathrm{out}}_n)\left(\begin{array}{c}
		\alpha^{\mathrm{out}}_n \\ 1
	\end{array}\right),
\end{equation}
where
$$
  \alpha^{\mathrm{in}}_n=\frac{\Delta_n(k^{\mathrm{in}}_{n})}{E+i\sqrt{|\Delta_n(k^{\mathrm{in}}_{n})|^2-E^2}},\quad 
  \alpha^{\mathrm{out}}_n=\frac{\Delta_n(k^{\mathrm{out}}_{n})}{E-i\sqrt{|\Delta_n(k^{\mathrm{out}}_{n})|^2-E^2}}.
$$
Following Ref. \onlinecite{Shel-bc}, the boundary condition has the form of a linear relation between the ``in'' and ``out'' envelope wave functions at $x=0$:
\begin{equation}
\label{Shelankov-bc-general}
  \phi(k^{\mathrm{out}}_n)=\sum_{m=1}^N \left( \begin{array}{cc}
          S_{nm} & 0 \\
          0 & \tilde S_{nm} 
          \end{array} \right)
          \phi(k^{\mathrm{in}}_{m}).
\end{equation}
The coefficients here form two $N\times N$ unitary matrices $\hat S$ and $\hat{\tilde S}$, which describe the boundary scattering of electrons and holes, respectively, at the Fermi level in the normal state.\cite{FHAB11} 

The relation (\ref{hole-Hamiltonian}) between the electron and hole Hamiltonians implies a certain relation between $\hat S$ and $\hat{\tilde S}$. 
The Fermi-level electron scattering matrix is found by solving the single-particle Schr\"odinger equation $\hat\varepsilon|\Psi\rangle=0$, where $\hat\varepsilon$ includes both
the bulk and boundary contributions. Asymptotically, \textit{i.e.}, inside the wire and sufficiently far from the boundary region, we have
$|\Psi\rangle=\sum_n(A_n|k^{\mathrm{in}}_{n},n\rangle+B_n|k^{\mathrm{out}}_{n},n\rangle)$. The $S$ matrix is defined by the relations
$$
  B_n=\sum_{nm}S_{nm}A_m.
$$
Similarly, the Fermi-level hole scattering matrix is found by diagonalizing the hole Hamiltonian $\hat{\tilde\varepsilon}$, which also includes the bulk and boundary terms.
It is easy to see that the state $|\tilde\Psi\rangle={\cal A}^{-1}|\Psi\rangle$ satisfies $\hat{\tilde\varepsilon}|\tilde\Psi\rangle=0$, see Eq. (\ref{hole-Hamiltonian}). Since 
$\hat{\tilde\varepsilon}=\hat\varepsilon$ in the bulk of the wire, we have $|\tilde\Psi\rangle=\sum_n(\tilde A_n|k^{\mathrm{in}}_{n},n\rangle+\tilde B_n|k^{\mathrm{out}}_{n},n\rangle)$ away from the boundary. By analogy with
Eq. (\ref{t_n-def}), one can write ${\cal A}^{-1}|k\rangle=\tau(k)|-k\rangle$, where $\tau(k)$ is a phase factor (here we omit the band indices, for brevity). It is easy to show that $\tau(k)=t(-k)$. 
Using $k^{\mathrm{out}}_{n}=-k^{\mathrm{in}}_{n}$, we have $\tilde A_n=B_n^*\tau_n(k^{\mathrm{out}}_{n})=B_n^*t_n(k^{\mathrm{in}}_{n})$ and $\tilde B_n=A_n^*\tau_n(k^{\mathrm{in}}_{n})=A_n^*t_n(k^{\mathrm{out}}_{n})$.
For the hole scattering matrix, which is defined by
$$
  \tilde B_n=\sum_{nm}\tilde S_{nm}\tilde A_m,
$$
we obtain the following expression:
\begin{equation}
\label{S-tilde S-general}
  \tilde S_{nm}=t_m^*(k^{\mathrm{in}}_{m})S_{mn}t_n(k^{\mathrm{out}}_{n}).
\end{equation}
Choosing the phases of the Bloch states in such a way that $t_n(k^{\mathrm{out}}_{n})=1$ for all bands and using the fact that $t_n$ is even (odd) in momentum for ${\cal A}^2=1$ (${\cal A}^2=-1$), we finally arrive at
\begin{equation}
\label{S-tilde S-relation}
  \hat{\tilde S}=\pm\hat S^\top,\quad \mathrm{for\ }{\cal A}^2=\pm 1.
\end{equation}
In general, the scattering matrix should be regarded as a phenomenological input. In some cases, if a treatable microscopic model of the band structure and the boundary potential is available,
then the scattering matrix can be calculated analytically, see the examples in Sec. \ref{sec: Rashba wire} below and in Ref. \onlinecite{SM16}. 

In the particular case of an ${\cal A}$-invariant normal state, when both the bulk of the wire and the boundary have the same antiunitary symmetry, we have $\hat{\tilde\varepsilon}=\hat\varepsilon$ everywhere and the 
electron-like and hole-like quasiparticles scatter in the same way. Therefore, $\hat{\tilde S}=\hat S$ and Eq. (\ref{Shelankov-bc-general}) becomes
\begin{equation}
\label{Shelankov-bc}
  \phi(k^{\mathrm{out}}_n)=\sum_{m=1}^N S_{nm}\phi(k^{\mathrm{in}}_{m}).
\end{equation}
In the ${\cal A}$-invariant case, Eq. (\ref{S-tilde S-relation}) becomes a constraint on the $S$ matrix: $\hat S^\top=\pm\hat S$ for ${\cal A}^2=\pm 1$. 

In general, if the boundary does not have the symmetry ${\cal A}$ of the bulk of the wire, then the $S$ and $\tilde S$ matrices are neither symmetric nor antisymmetric, but are still related by Eq. (\ref{S-tilde S-relation}). 
Inserting the envelope functions (\ref{phi-in-out}) into the boundary conditions (\ref{Shelankov-bc-general}), 
we obtain a system of $2N$ linear equations for $C(k^{\mathrm{in}}_n)$ and $C(k^{\mathrm{out}}_n)$. It has a nontrivial solution if 
\begin{equation}
\label{ABS-energy-eq-gen}
  \det\hat W(E)=0,
\end{equation}
where $\hat W(E)=\hat{\tilde S}-\diag(\alpha^{\mathrm{out},*}_1,...,\alpha^{\mathrm{out},*}_N)\hat S\diag(\alpha^{\mathrm{in}}_1,...,\alpha^{\mathrm{in}}_N)$.
Using Eqs. (\ref{Type-II-OP}), (\ref{Type-III-OP}), and (\ref{S-tilde S-relation}), this can be represented in the following form:
$$
  \hat W(E)=\pm\left[\hat S^\top-\hat D^\dagger(E)\hat S\hat D(-E)\right],\quad \mathrm{for\ }{\cal A}^2=\pm 1,
$$
where $\hat D(E)=\diag[e^{i\theta_1(E)},...,e^{i\theta_N(E)}]$ and we introduced the notation $\alpha^{\mathrm{out}}_n(E)=e^{i\theta_n(E)}$. Finally, observing that $\hat W^\top(E)=-\hat D(-E)\hat W(-E)\hat D^\dagger(E)$, 
we obtain:
\begin{equation}
\label{WE-E}
  \det\hat W(-E)=(-1)^N e^{i\sum_n[\theta_n(E)-\theta_n(-E)]}\det\hat W(E).
\end{equation}
One can draw two important conclusions from this last equation. First, the ABSs with nonzero energies come in pairs: if $E$ is the energy of an ABS localized near one end of the wire (near $x=0$), 
then there is another ABS localized near the same end, with energy $-E$. Second, since
$\det\hat W(0)=(-1)^N\det\hat W(0)$, there is at least one zero-energy ABS if $N$ is odd, in agreement with the argument presented in Ref. \onlinecite{Kit01}. There is another zero-energy ABS localized near 
the other, infinitely remote, end of the wire. These two solutions can be interpreted as two Majorana modes, or two ``halves'' of one highly nonlocal usual (\textit{i.e.}, complex) fermionic mode.  
In a wire of a finite length $l$, the zero-energy ABSs hybridize, producing two states with exponentially small nonzero energies $\pm{\cal O}(e^{-\Delta l/v_F})\Delta$. 

We see that generically there is no zero-energy ABS if the number of bands is even, \textit{i.e.}, in all Type II wires, as well as in Type III wires with an even $N$. In these systems, the zero-energy states may exist, 
but they are unstable against any ${\cal A}$ symmetry-breaking perturbation. 
In contrast, if the number of bands is odd, which is only possible in Type III wires, then there always exists at least one zero-energy Majorana mode near the end of a half-infinite wire,
regardless of the relative phases of $\eta_n$ and the details of the boundary scattering.
This can also be understood using the language of the tenfold symmetry classification (Ref. \onlinecite{tenfold-way}). 
If the ${\cal A}$ symmetry of a Type III wire is broken (by the superconducting order parameter and/or the boundaries), then the whole system is in Class D, characterized in 1D by a 
$\mathbb{Z}_2$ invariant $(-1)^N$, and is therefore topologically nontrivial if $N$ is odd. If the ${\cal A}$ symmetry is not broken, then the system is in Class BDI, which is characterized in 1D by a 
$\mathbb{Z}$ invariant,\cite{Z-invariant} whose simple explicit form was found in Ref. \onlinecite{Sam17-AoP}, see Eq. (\ref{N_MF-BDI}) below.

\section{Rashba wire in a magnetic field}
\label{sec: Rashba wire}

As an example of the theory developed in the previous section, let us consider a half-infinite wire in which the antisymmetric SOC is described by the Rashba model\cite{Rashba-model} and TR symmetry is broken by 
a magnetic field $H$ along the $z$ axis. The Hamiltonian of the wire at $x>0$ has the following form:
\begin{equation}
\label{Rashba-Hamiltonian}
  \hat\varepsilon=\frac{\hat k_x^2}{2m^*}-\epsilon_F+\bm{\Gamma}(\hat k_x)\hat{\bm\sigma},
\end{equation}
where the first term describes a single spin-degenerate 1D band in the effective mass approximation, $\epsilon_F=k_F^2/2m^*$ is the Fermi energy, and $\hat k_x=-i\nabla_x$. The last term, where
$\bm{\Gamma}(k)=ak\hat{\bm y}+\mu_BH\hat{\bm z}$, $\hat{\bm{\sigma}}$ are the Pauli matrices, and $\mu_B$ is the Bohr magneton, combines the 
Rashba SOC and the Zeeman energy. The boundary is described by an infinitely high wall at $x=0$. The electron spectrum in the bulk of the wire (at $x>0$) consists of two nondegenerate bands
\begin{equation}
\label{equation}
  \xi_\lambda(k)=\frac{k^2-k_F^2}{2m^*}+\lambda\sqrt{a^2k^2+\mu_B^2H^2},
\end{equation}
labelled by the helicity $\lambda=\pm$. Depending on the magnetic field strength, the number of bands crossing the Fermi level is either two, at $\mu_BH<\epsilon_F$, or one, at $\mu_BH>\epsilon_F$. 

According to Sec. \ref{sec: symmetry}, the Hamiltonian (\ref{Rashba-Hamiltonian}) describes a Type III wire with the magnetic class 
$\mathbb{G}_M=\mathbf{D}_y(E)$ or $\mathbf{V}(\mathbf{C}_2)$. The antiunitary symmetry ${\cal A}=K\sigma_y$ is preserved by the boundary, but may be broken by the superconducting order parameter.
Superconductivity in this model has $p$-wave symmetry, see Eq. (\ref{Type-III-OP}), corresponding to a two-band or one-band continuous version of the Kitaev chain.

\subsection{$N=2$}
\label{sec: N2}

\begin{figure}
\includegraphics[width=8cm]{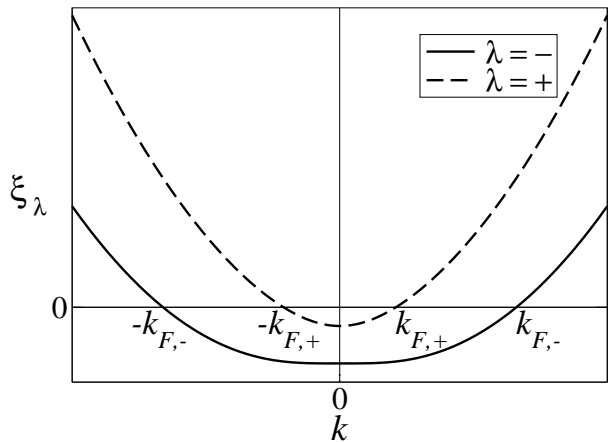}
\caption{The helicity bands in the case $N=2$ (weak magnetic field).}
\label{fig: N2}
\end{figure}

The band spectrum of the wire in the weak field limit is shown in Fig. \ref{fig: N2}. The general wave function of the Fermi-level quasiparticles at $x>0$ is a superposition of two incident and 
two reflected Bloch waves:
\begin{equation}
\label{Psi-general-N2}
 \Psi(x)=A_-\langle x|k^{\mathrm{in}}_{-},-\rangle+A_+\langle x|k^{\mathrm{in}}_{+},+\rangle+
  B_-\langle x|k^{\mathrm{out}}_{-},-\rangle+B_+\langle x|k^{\mathrm{out}}_{+},+\rangle,
\end{equation}
where $k^{\mathrm{out}}_{\lambda}=k_{F,\lambda}$ and $k^{\mathrm{in}}_{\lambda}=-k_{F,\lambda}$. Using the parameterization $\bm{\Gamma}(k)=|\bm{\Gamma}(k)|(\hat{\bm y}\sin\theta_k+\hat{\bm z}\cos\theta_k)$, with
$$
  \theta_k=\arctan\left(\frac{ak}{\mu_BH}\right),
$$
the eigenstates of the Hamiltonian (\ref{Rashba-Hamiltonian}) have the following form:
\begin{eqnarray}
\label{states-in-out}
  \langle x|k,+\rangle=\frac{i}{\sqrt{|v_{+}(k)|}}\left(\begin{array}{c}
                                       \cos(\theta_k/2) \\ i\sin(\theta_k/2)
                                      \end{array}\right)e^{ikx},\quad
  \langle x|k,-\rangle=\frac{i}{\sqrt{|v_{-}(k)|}}\left(\begin{array}{c}
                                       i\sin(\theta_k/2) \\ \cos(\theta_k/2)
                                      \end{array}\right)e^{ikx},
\end{eqnarray}
where $v_\lambda(k)=\partial\xi_\lambda/\partial k$. We use the normalization in which the magnitude of the probability current carried 
by each of the plane-wave states (\ref{states-in-out}) is equal to one.

The phases of the band states are fixed as follows. According to Eq. (\ref{t_n-def}), we have ${\cal A}|k,\lambda\rangle=t_\lambda(k)|-k,\lambda\rangle$. It is easy to check that 
the action of the antiunitary symmetry ${\cal A}=K\sigma_y$ on spin-$1/2$ wave functions in 1D is given by ${\cal A}\Psi(x)=-\Psi^*(x)$. Using the fact that $\theta_{-k}=-\theta_k$ and choosing $t_\lambda(k)=1$ for all $k$, 
which is the simplest form compatible with the requirement that $t_\lambda(-k)=t_\lambda(k)$, we arrive at Eq. (\ref{states-in-out}).

From the boundary condition for the wave function (\ref{Psi-general-N2}) at an infinitely high wall,
\begin{equation}
\label{boundary-condition}
  \Psi(x=0)=0,
\end{equation}
we obtain two linear relations between the amplitudes of the reflected ($B_\pm$) and incident ($A_\pm$) waves. These relations can be written in the form
$$
  \left(\begin{array}{c}
  B_- \\ B_+
  \end{array}\right)=
  \hat S
  \left(\begin{array}{c}
  A_- \\ A_+
  \end{array}\right),
$$
where 
\begin{eqnarray}
\label{S-matrix-N2}
  \hat S =-\frac{1}{\cos\left(\frac{\theta_+-\theta_-}{2}\right)}
	 \left(\begin{array}{cc}
         \cos\left(\frac{\theta_++\theta_-}{2}\right) & -i\sqrt{\frac{v_{F,-}}{v_{F,+}}}\sin\theta_+ \\
         -i\sqrt{\frac{v_{F,+}}{v_{F,-}}}\sin\theta_- & \cos\left(\frac{\theta_++\theta_-}{2}\right)
         \end{array}\right).
\end{eqnarray}
is the boundary scattering matrix of the Fermi-level quasiparticles. Here $\theta_\lambda=\theta(k_{F,\lambda})$ and $v_{F,\lambda}=v_\lambda(k_{F,\lambda})$. Since both the wire and the boundary are 
${\cal A}$-invariant and ${\cal A}^2=+1$, one should have $\hat{\tilde S}=\hat S^\top=\hat S$, see Sec. \ref{sec: ABS}. It is straighforward to show that $v_{F,+}/\sin\theta_+=v_{F,-}/\sin\theta_-$, which means that the 
$S$ matrix (\ref{S-matrix-N2}) is indeed symmetric and unitary. 

Let us now take into account superconducting pairing. According to Eq. (\ref{Type-III-OP}), the gap functions in the helicity bands are given by $\Delta_\lambda(k)=i\eta_\lambda\varphi_\lambda(k)$, 
where $\varphi_\lambda(k)\propto k$ are $p$-wave basis functions. The order parameter components $\eta_\pm$ can be generally written as 
$$
  \eta_-=|\eta_-|,\quad \eta_+=|\eta_+|e^{i\chi}.
$$
If the phase difference $\chi$ is equal to $0$ or $\pi$, then the order parameter is real and the antiunitary symmetry ${\cal A}$ of the normal state is preserved in the superconducting state. At any other value of 
$\chi$, the superconducting state is not ${\cal A}$ invariant. 

Substituting the scattering matrix (\ref{S-matrix-N2}) in the ABS energy equation (\ref{ABS-energy-eq-gen}), we obtain:
\begin{equation}
\label{ABS-eq-Rashba-wire-N2}
  \sqrt{(|\eta_-|^2-E^2)(|\eta_+|^2-E^2)}-|\eta_-||\eta_+|\cos\chi=RE^2,
\end{equation} 
where
$$
  R=\frac{1+\cos\theta_-\cos\theta_+}{\sin\theta_-\sin\theta_+}.
$$
Therefore, the ABS spectrum consists of symmetric pairs $\pm|E|$, and the zero-energy states, if they exist, are twofold degenerate. It follows from Eq. (\ref{ABS-eq-Rashba-wire-N2}) that the zero-energy ABS (the Majorana modes) 
are present only if $\chi=0$, which agrees with the topological arguments. Indeed, at $\chi=0$ or $\pi$ the wire, the boundary, and the superconducting order parameter are all ${\cal A}$ invariant, 
so that the system is in the symmetry class BDI and the relevant $\mathbb{Z}$ topological invariant, which gives the number of the MFs, is 
\begin{equation}
\label{N_MF-BDI}
  N_{\mathrm{MF}}=|N_+-N_-|,
\end{equation}
where $N_+$ ($N_-$) is the number of bands with positive (negative) order parameter (Ref. \onlinecite{Sam17-AoP}). For $\chi=0$ this expression yields $N_{\mathrm{MF}}=2$, while for $\chi=\pi$ we have 
$N_{\mathrm{MF}}=0$.  

According to Eq. (\ref{ABS-eq-Rashba-wire-N2}), the two zero-energy ABSs are unstable against a small fluctuation of the relative phase: if $\chi=\delta\chi$, where $|\delta\chi|\ll 1$, then
$$
  E=\pm|\delta\chi|\sqrt{\frac{|\eta_-|^2|\eta_+|^2}{|\eta_-|^2+|\eta_+|^2+2R|\eta_-||\eta_+|}}.
$$
This is again consistent with the topological arguments. At $\delta\chi\neq 0$, the ${\cal A}$ symmetry is broken in the superconducting state, changing the symmetry class from BDI to D. The $\mathbb{Z}_2$
topological invariant is $(-1)^N$ and the MF number is given by
\begin{equation}
\label{N_MF-D}
  N_{\mathrm{MF}}=\frac{1-(-1)^N}{2}.
\end{equation}
Therefore, the system with $N=2$ is topologically trivial and does not have MFs.

\subsection{$N=1$}
\label{sec: N1}

\begin{figure}
\includegraphics[width=8cm]{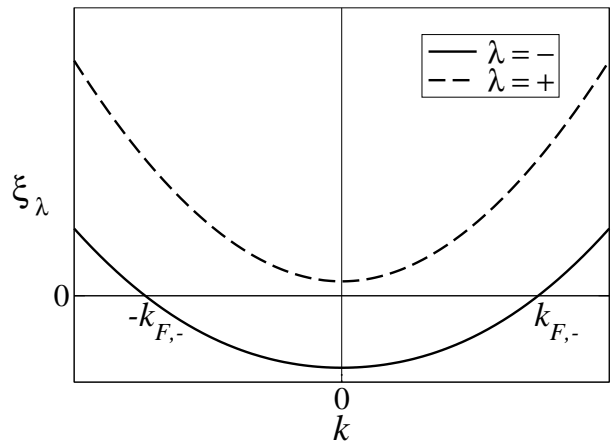}
\caption{The helicity bands in the case $N=1$ (strong magnetic field).}
\label{fig: N1}
\end{figure}

The band spectrum of the wire in the strong field limit is shown in Fig. \ref{fig: N1}. In this case, the states corresponding to the minority band ($\lambda=+$) are described by evanescent waves in the bulk of the wire. 
Although the positive helicity states do not participate in the superconducting pairing, one has to take them into account when calculating the boundary scattering matrix, in order to satisfy the boundary condition. 
The wave function of the Fermi-level quasiparticles at $x>0$ has the form
\begin{equation}
\label{Psi-general-N1}
 \Psi(x)=A_-\langle x|k^{\mathrm{in}}_{-},-\rangle+B_-\langle x|k^{\mathrm{out}}_{-},-\rangle+C\left(\begin{array}{c}
               1 \\ r
               \end{array}\right)e^{-\kappa x}.
\end{equation}
The first two terms here are the propagating waves in the majority band, see Eq. (\ref{states-in-out}), while in the last term we have
$$
  r=-\frac{a\kappa}{\mu_BH+\sqrt{\mu_B^2H^2-a^2\kappa^2}},\quad \kappa=k_F\left(\sqrt{p_1^2+p_2^2+\frac{p_1^4}{4}}-1-\frac{p_1^2}{2}\right)^{1/2},
$$
where $p_1=ak_F/\epsilon_F$ and $p_2=\mu_BH/\epsilon_F$.

Substituting Eq. (\ref{Psi-general-N1}) in the boundary condition (\ref{boundary-condition}), we obtain the following expression for the only element of the $S$-matrix:
\begin{equation}
\label{S-matrix-N1}
  S_{--}=\frac{B_-}{A_-}=-\frac{1+ir\tan(\theta_-/2)}{1-ir\tan(\theta_-/2)}.
\end{equation}
It is easy to see that $|S_{--}|=1$, in agreement with the particle number conservation, which requires $|B_-|^2=|A_-|^2$.

The superconducting order parameter has one component $\eta_-$ and the ABS energy equation (\ref{ABS-energy-eq-gen}) takes the form
\begin{equation}
\label{ABS-eq-Rashba-wire-N1}
  1+\frac{|\eta_-|^2}{(E+i\sqrt{|\eta_-|^2-E^2})^2}=0.
\end{equation}
The only solution is $E=0$, which corresponds to a nondegenerate zero-energy ABS and is interpreted as a single MF. 

A remarkable feature of Eq. (\ref{ABS-eq-Rashba-wire-N1}) is that it does not contain $S_{--}$ and is therefore completely insensitive to the details of the boundary scattering. 
This means that the Majorana state is always present in a single-band wire, in agreement with the general analysis in Sec. \ref{sec: ABS}.
The symmetry class is either BDI (for an ${\cal A}$-invariant boundary, such as an infinitely high wall) or D (for an ${\cal A}$ symmetry-breaking boundary). According to Eqs. (\ref{N_MF-BDI}) and (\ref{N_MF-D}),
$N_{\mathrm{MF}}=1$ in both cases.

\section{Conclusion}
\label{sec: conclusion}

To summarize, we calculated analytically the ABS spectrum in a half-infinite superconducting wire with an arbitrary number of bands, without resorting to any particular microscopic model of the band structure. 
We have proved that there is one zero-energy MF at the end of the wire if (i) the normal state of the wire has an antiunitary symmetry ${\cal A}$, satisfying ${\cal A}^2=1$, and (ii) the number of bands 
crossing the chemical potential is odd. This MF is still present even when the superconducting order parameter and/or the boundary scattering are not ${\cal A}$-invariant. The general analysis is illustrated using as 
an example a half-infinite Rashba wire in a perpendicular magnetic field.

\acknowledgments

This work was supported by a Discovery Grant from the Natural Sciences and Engineering Research Council of Canada.

\end{document}